\documentclass[12pt,a4paper]{article}

\usepackage[utf8]{inputenc}
\usepackage{fullpage}
\usepackage{hyperref}
\usepackage{amssymb}
\usepackage{amsthm}
\usepackage{graphicx}
\usepackage[outdir=./]{epstopdf}
\usepackage{color}
\usepackage{enumitem}
\usepackage{empheq}
\usepackage{cite}
\usepackage{slashed}
\usepackage{amsmath}
\usepackage{braket}
\usepackage{simpler-wick}
\usepackage{simplewick}

\numberwithin{equation}{section}
\usepackage{pgfplots}
\ifdim\overfullrule>0pt 
\usepackage{environ}

\NewEnviron{tikzpicture}{%
	\begin{pgfpicture}
		\pgfpathrectanglecorners{\pgfpointorigin}{\pgfpoint{3cm}{3cm}}%
		\pgfusepath{stroke}\end{pgfpicture}%
}
\fi

\renewcommand{\Re}{\operatorname{Re}}

\newcommand{\e}{\epsilon}

\def\Tr{\mathop{\rm Tr}}

\usepackage{cleveref}
\crefname{section}{Section}{Sections}
\crefname{appendix}{Appendix}{Appendices}

\begin{document}

	\titlepage
	
	\begin{flushright}
		MS-TP-21-40
	\end{flushright}

	\vspace*{1.2cm}
	
	\begin{center}
		{\Large \bf Soft photon bremsstrahlung
			at 
			next-to-leading power}
		
		\vspace*{1.5cm} \textsc {D. Bonocore, A. Kulesza} \\
		
		\vspace*{1cm}

		Institut f\"{u}r Theoretische Physik, Westf\"{a}lische
		Wilhelms-Universit\"{a}t M\"{u}nster, Wilhelm-Klemm-Stra\ss e 9,
		D-48149 M\"{u}nster, Germany\\

	\end{center}

	\vspace*{7mm}	
	\begin{abstract}
		\noindent
		The emission of soft radiation provides a fundamental probe of the 
	consistency of the underlying quantum field theory. Correspondingly, the 
	measurement of the soft photon bremsstrahlung, such as 
	the one proposed with the planned future upgrade of the ALICE experiment at 
	the LHC, is of great interest.   
	In this letter we 
	explore the possibility to implement analytic techniques that have been 
	recently 
	developed for soft gluon resummation at Next-to-Leading-Power (NLP) in the 
	context of the soft 
	photon spectrum. 
	We provide a formula for the differential cross-section with shifted 
	kinematics 
	that is 
	particularly 
	suitable for numerical implementations. We also discuss the impact of loop  
	corrections to Low's theorem due to radiative jet functions.
	\end{abstract}

\newpage


\section{Introduction}
\label{sec:intro}


There are plans for a new multipurpose detector at the Large Hadron Collider 
(LHC) as a follow-up to the present ALICE experiment \cite{Adamova:2019vkf}. In 
the rich physics program that would be enabled by this upgrade, the possibility 
to measure ultra soft photons at very low transverse momentum is particularly 
attractive. In fact, 
	measurements of soft photon spectra have been undertaken by many 
experiments 
over the last couple of decades, motivated by a clean theoretical picture of 
photon emissions. Yet, to this day the results of the measurements remain 
not understood, as yields of soft photons produced together with hadrons 
show significant excess above theoretical predictions
\cite{Chliapnikov:1984ed, EHSNA22:1991sdp, 
	SOPHIEWA83:1992czx, WA91:1997cnv, Belogianni:2002ic, Belogianni:2002ib, 
	DELPHI:2005yew, DELPHI:2007nmh, 
	DELPHI:2010cit, Wong:2014ila}.
In the light of this 
outstanding discrepancies, and future measurements planned at the LHC, one 
needs to scrutinize the theoretical predictions used to compare 
with data.

The 
theoretical foundations of soft boson emissions have a 
long history that have continued to attract attention until recent days 
\cite{Gell-Mann:1954wra, Low:1958sn, Weinberg:1965nx, Burnett:1967km, 
	Bell:1969yw, Cachazo:2014fwa, Strominger:2017zoo}. 
Surprisingly, the theoretical description of soft photon emission 
spectra employed so far relies only on the leading-power 
(LP) eikonal approximation,
 which corresponds to 
terms of order $1/\omega_k$ in Low's soft theorem 
\cite{Low:1958sn}, where $\omega_k$ is the energy of the  
photon 
with soft momentum $k^{\mu}$.
The corresponding 
soft photon bremsstrahlung cross section at LP reads

\begin{align}
\frac{d\sigma_{\text{LP}}}{d^3k} & =\frac{\alpha}{(2\pi)^2} 
\frac{ 1}{\omega_k} \int d^3p_3\dots\int 
d^3p_{n}
\left(\sum_{i,j=1}^{n}-\eta_i\eta_j
\frac{p_i\cdot p_j}{(p_i\cdot k)( p_j\cdot k)}\right)
d\sigma_H(p_1,\dots, p_n)~,
\label{crossLP}
\end{align}
where $\alpha$ is the electromagnetic constant and $d\sigma_H$ denotes the 
differential 
non-radiative  
cross-section 
depending on the $n$ hard momenta  
$p_i^{\mu}$ of the  
incoming and outgoing charged particles. The sign of $\eta=\pm1$ is equal 
(opposite) to the sign of 
the electric charge of the final (initial) states.

It is one of the main purposes of this letter to show that
possible improvements in the description of the photon emission spectra can be 
achieved by employing methods which have been already developed in the QCD 
resummation program.
In that context,
the soft boson is 
an 
\emph{unobserved} 
gluon giving rise to large logarithms in the cross-section that need to be 
resummed to all orders 
in perturbation theory \cite{Sterman:1986aj, Catani:1989ne}. In this regard, 
there has been a great deal of 
interest in the recent years in extending this framework to 
Next-to-Leading Power 
(NLP), in order to control the effects of power 
suppressed emissions of soft gluons and quarks \cite{Laenen:2008gt, 
	Bonocore:2020xuj, DelDuca:2017twk, vanBeekveld:2019cks, 
	vanBeekveld:2019prq, 
	Bahjat-Abbas:2019fqa, vanBeekveld:2021hhv, vanBeekveld:2021mxn, 
	Moult:2017jsg, Moult:2018jjd, Ebert:2018lzn, Moult:2019mog, 
	Moult:2019vou, Beneke:2017ztn, Beneke:2018gvs, Beneke:2018rbh, 
	Beneke:2019oqx,  Boughezal:2018mvf, Liu:2020tzd, Liu:2020wbn, Liu:2021mac, 
	Broggio:2021fnr, Inglis-Whalen:2021bea, Ajjath:2020sjk, Ajjath:2020lwb}. 
In this letter we explore the possibility to implement 
these recently developed NLP techniques to the \emph{observed} soft photon 
emissions.
In particular, we discuss two distinct corrections to \cref{crossLP}
to which NLP effects give rise.

The first of these corrections are of order $(\omega_k)^0\sim 1$ and have first 
been analyzed by Low for scalar emitters, and later extended by Burnett and 
Kroll to 
spinor emitters \cite{Burnett:1967km, Bell:1969yw}. In this letter,
building on similar results derived in QCD for processes with 
colorless final 
states 
\cite{DelDuca:2017twk}, we 
propose a form of the Low-Burnett-Kroll (LBK) formula in terms of 
kinematical shifts of external particle momenta. This formulation is 
particularly suitable for numerical implementations. In 
doing so, we clarify also an aspect that has been recently 
addressed in \cite{Lebiedowicz:2021byo} regarding the validity of 
$(\omega_k)^0$ 
corrections when the non-radiative 
amplitude is expressed in terms of non-physical momenta that violate momentum 
conservation.

 Furthermore, the method of shifting external particle momenta has 
proven invaluable while deriving soft gluon resummed expressions in Mellin 
space 
at NLP for processes with colorless final states. It is therefore justified to 
presume that generalizing the shifting procedure for 
processes with arbitrary number of charged external legs will be very useful 
in the extension of QCD resummation at NLP for processes with more than two  
colored 
external legs.

The second kind of corrections is due to an interplay between the 
soft photon emission and QCD loop effects 
\cite{Bern:2014oka, 
	He:2014bga, 
	Larkoski:2014bxa, Bonocore:2014wua}. 
They have a logarithmic form in photon momentum and therefore could 
significantly 
increase yields of soft photon emissions. In particular, as they are due to QCD 
virtual contributions, they would only appear in calculations involving colored 
external lines. Hence, they might help to 
explain the observed discrepancies 
\cite{Chliapnikov:1984ed, EHSNA22:1991sdp, 
	SOPHIEWA83:1992czx, WA91:1997cnv, Belogianni:2002ic, Belogianni:2002ib, 
	DELPHI:2005yew, DELPHI:2007nmh, 
	DELPHI:2010cit, Wong:2014ila} in the 
production rates.  
Although in 
principle one 
could consider loop corrections from radiative jets both in QED and QCD, it 
is clear 
that the contribution from the latter is larger because of the strong  
coupling constant. 
Therefore, one expects the effect of these 
corrections to be more important when soft photons are  
emitted from hadrons, rather than leptons.

More specifically, virtual collinear contributions modify the structure of 
the LBK theorem,\footnote{An extension of the LBK theorem 
	accounting for soft virtual 
	contributions has been recently discussed in \cite{Engel:2021ccn}.}  
though these effects are absent if 
the soft photon 
energy $\omega_k\ll\frac{m^2}{Q}$, where $m$ is 
the mass of the lightest charged particle and $Q$ is the typical energy of 
the 
process. However, this 
condition is not fulfilled for massless particles, which are 
needed in the standard partonic framework of scattering amplitudes.
Similarly, it would not be fulfilled for high-energy processes with leptons in 
the final state. 
In order to extend the soft theorem at the
loop level to the larger region 
$\frac{m^2}{Q}\le \omega_k\sim m$ (which includes the massless limit)
one needs to take into account so-called  \emph{radiative jets} 
\cite{DelDuca:1990gz, Bonocore:2015esa, Bonocore:2016awd, 
	Gervais:2017yxv, Laenen:2020nrt}. 
We point out that 
they
give rise to corrections of the form  
 $\log(\mu^2/p_i\cdot k)$ to 
\cref{crossLP}, where $\mu$ is the renormalization scale.  
Therefore, they are enhanced for very small photon 
energies, and in particular for soft photons with very low transverse 
momentum.  
At the amplitude level, our calculation with radiative jets clarifies the 
massless limit of 
logarithmic 
corrections to soft photon theorems 
derived with 
a somewhat 
unconventional regularization of 
infrared divergences in \cite{Laddha:2018myi, Sahoo:2018lxl}.

The structure of this letter is as follows. We begin in \cref{sec:LBK} with a 
review of the LBK theorem at NLP. In \cref{sec:shifts} we discuss how to 
implement the LBK theorem with kinematical shifts. In \cref{sec:loop} we 
discuss the impact of QCD corrections with radiative jet functions. Finally, 
we discuss our results in \cref{sec:concl}.


\section{Review of the LBK theorem}
\label{sec:LBK}

\begin{figure}
	\centering
	\includegraphics[width=50mm]{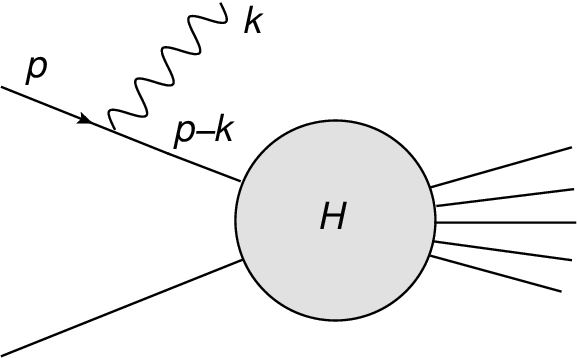}
	\qquad 
	\includegraphics[width=50mm]{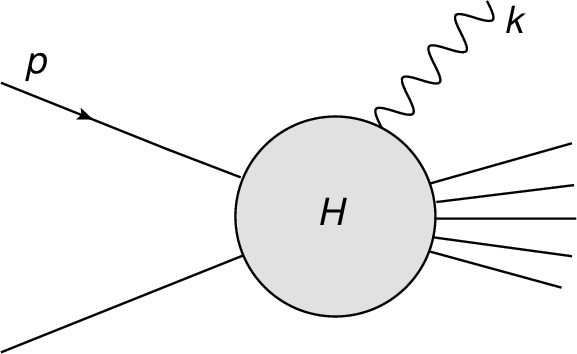}
	\caption{Diagrammatic representation of a soft photon emission from an 
		external line (left) and from an 
		internal line (right) in a scattering amplitude. The blob ${\cal 
			H}$ represents a generic subdiagram sensitive to the unspecified 
			hard 
		dynamics.}
	\label{fig:internal}
\end{figure}

In this section we briefly review the original theorem by Low, Burnett and 
Kroll in the 
language used in current literature.  
Without loss of generality, we consider the emission of a soft photon of 
momentum $k$ in a scattering 
amplitude ${\cal A}(p_1,p_2, k)$, where for simplicity the only charged 
particles are the incoming
particle-antiparticle pair of spin $1/2$, mass $m$ and momenta $p_1$ and 
$p_2$. The extension to more general processes is straightforward.

We start at the amplitude level. 
Diagrammatically, the external lines of these amplitudes are connected  
via a hard subdiagram ${\cal H}(p_1,p_2)$, representing an unspecified 
hard dynamics. 
As depicted in \cref{fig:internal}, the photon can couple either to one of the 
external line or to the internal hard blob ${\cal H}$. We define the 
corresponding external and internal diagrams
after stripping off the photon polarization vector $\epsilon_{\mu}(k)$ as 
${\cal A}_{\text{ext}}^{\mu}$ and 
${\cal A}_{\text{int}}^{\mu}$, 
respectively. Let us compute them separately.

The external emission from the particle of momentum $p_1$
reads\footnote{For simplicity, we set the electric charge $e$ to unity.}
\begin{align}
{\cal A}_{\text{ext},1}^{\mu}
&=-\eta_1\bar v(p_2){\cal H}(p_1-k,p_2)
\frac{\slashed{p}_1-\slashed{k}+m}{(p_1-k)^2-m^2}\gamma^{\mu}u(p_1)\notag \\
&=\eta_1\bar v(p_2){\cal H}(p_1,p_2)
\left(
\frac{p_1^{\mu}}{p_1\cdot k}-\frac{k^{\mu}}{2p_1\cdot k}
+\frac{k^2 p_1^{\mu}}{2(p_1\cdot k)^2}
-\frac{ik_{\nu}S^{\mu\nu}}{p_1\cdot 
	k}
\right) u(p_1) \notag \\
&\quad -\eta_1\frac{p^{\mu}_1}{p_1\cdot k}k^{\nu}
\bar v(p_2)\frac{\partial {\cal H}(p_1,p_2)}{\partial 
	p_1^{\nu}} u(p_1)\,\, + \,\,{\cal O}(k)~.
\label{ext}
\end{align}
where we introduced the Lorentz generator 
$S^{\mu\nu}=\frac{i}{4}[\gamma^{\mu},\gamma^{\nu}]$. In \cref{ext} we expanded
the amplitude at NLP 
in the soft photon momentum $k$ and  exploited the functional dependence of 
${\cal H}$ to set
\begin{align}
\frac{\partial {\cal H}(p_1-k,p_2)}{\partial 
	k^{\nu}}\Big|_{k=0}
=-\frac{\partial {\cal H}(p_1,p_2)}{\partial 
	p^{\nu}_1}~.
\end{align}
The first term in the second line of \cref{ext} is of order $1/k$ and gives rise
to the LP cross-section in \cref{crossLP}. The remaining terms are suppressed 
by a power of $k$ and therefore are NLP 
corrections. 
The emission for the incoming antiparticle
follows analogously. 

The internal emission diagram cannot be naively evaluated since we do not know 
how the photon couples with the short distance dynamics of the hard function. 
However, by exploiting the gauge invariance of the full amplitude, one can 
use the following Ward identity
\begin{align}
k_{\mu}\left(
\sum_{i=1}^2 {\cal A}_{\text{ext},i}^{\mu} + {\cal A}_{\text{int}}^{\mu}
\right)=0
\end{align}
to get
\begin{align}
{\cal A}_{\text{int}}^{\mu}&=
\sum_{i=1}^2 \eta_i\,
\bar v (p_2)
\frac{\partial {\cal H}(p_1,p_2)}{\partial 
	p_{i\,\mu}} u(p_1)~.
\label{int}
\end{align}
Here we assumed that the terms in ${\cal A}_{\text{int}}^{\mu}$ which are 
separately transverse to $k^{\mu}$ can be ignored at NLP. While this follows 
straightforwardly at 
tree-level from power counting, 
special care must be taken when loops of collinear lines are present   
\cite{Gervais:2017yxv}. 

One can then combine external and internal emissions. Enforcing the on-shell 
condition on the radiated photon yields the following LBK 
theorem at the amplitude level:
\begin{align}
{\cal A}^{\mu}(p_1,p_2,k)
&=\sum_{i=1}^2 \eta_i\,\frac{p_i^{\mu}}{p_i\cdot k}\,\bar v(p_2){\cal 
H}(p_1,p_2)
u(p_1) 
+\sum_{i=1}^2 \eta_i\,\bar v (p_2) 
G^{\mu\nu}_i\,\frac{\partial {\cal H}(p_1 ,p_2)
}{\partial p_i^{\nu}}
u(p_1)
\notag \\
&\quad + \eta_2\,\bar v (p_2) \frac{ik_{\nu}S^{\mu\nu}}{p_2\cdot k}{\cal H}(p_1 
,p_2) u(p_1)
- \eta_1\,\bar v (p_2) {\cal H}(p_1 ,p_2) \frac{ik_{\nu}S^{\mu\nu}}{p_1\cdot k} 
u(p_1)
~,
\label{ext+int}
\end{align}
where we defined 
\begin{align}
G^{\mu\nu}_i=g^{\mu\nu}-\frac{(2p_i-k)^{\mu}k^{\nu}}{2p_i\cdot k}
=g^{\mu\nu}-\frac{p_i^{\mu}k^{\nu}}{p_i\cdot k}+{\cal O}(k)~.
\label{Gtensor}
\end{align}
The first term on the r.h.s. of \cref{ext+int} corresponds to the LP soft 
theorem, while the 
other terms corresponds to NLP corrections.

It is common in the contemporary literature to introduce the orbital angular 
momentum 
generator 
$L^{\mu\nu}_i=i\left(p_i^{\mu}\frac{\partial}{\partial p_{i\,\nu}}
-p_i^{\nu}\frac{\partial}{\partial p_{i\,\mu}}\right)$. Indeed, 
by	observing that
$G^{\mu\nu}_i\,\frac{\partial}{\partial p_i^{\nu}}
=i\frac{k_{\nu}}{p_i\cdot k}L^{\mu\nu}_i$, one can rewrite \cref{ext+int} in 
terms of the total angular momentum 
$J^{\mu\nu}_i=S^{\mu\nu}_i+L^{\mu\nu}_i$	  
as
\begin{align}
\epsilon_{\mu}(k){\cal A}^{\mu}(p_1,p_2,k)
&=\left({\cal S}_{\text{LP}}+{\cal S}_{\text{NLP}}\right){\cal A}(p_1,p_2)~,
\label{next-to-soft}
\end{align}
where ${\cal A}(p_1,p_2)$ is the non-radiative amplitude and 
\begin{align}
{\cal S}_{\text{LP}}=\sum_{i=1}^2 \eta_i\,\frac{p_i\cdot \epsilon(k)}{p_i\cdot 
	k}~, \qquad
{\cal S}_{\text{NLP}}=\sum_{i=1}^2 
\eta_i\,\frac{ik_{\nu}J^{\mu\nu}_i\epsilon_{\mu}(k)}{p_i\cdot 
	k}~.
\label{SLP}
\end{align}
This factorized form of the theorem emphasizes that the NLP term is coupled to 
the total angular momentum of the hard emitter. 
One should be careful though since ${\cal S}_{\text{NLP}}$ in 
\cref{next-to-soft}
is not a multiplicative factor, but rather an operator acting on the hard 
function 
only, as made explicit in 
\cref{ext+int}.

After squaring \cref{ext+int}, summing over the polarizations and neglecting 
NNLP terms, we get
the LBK theorem for the unpolarized squared amplitude, which reads 
\begin{align}
|{\cal A}(p_1,p_2,k)|^2
&=\sum_{ij}(-\eta_i\eta_j)\frac{p_i\cdot p_j}{p_i\cdot k \,p_j\cdot k}|{\cal 
	A}(p_1,p_2)|^2 
+\sum_{ij}(-\eta_i\eta_j)\frac{p_{i\,\mu} }{p_i\cdot k }
G_j^{\mu\nu} \frac{\partial}{\partial p_j^{\nu}}|{\cal A}(p_1,p_2)|^2 ~.
\label{squared}
\end{align}
Eq.~(\ref{squared}) is valid both for scalar and spinning particles. This is 
straightforward in the former case, since after setting the spin 
generator to 
zero in \cref{ext+int} and replacing all spinor wave functions with unity, we 
are left only with the orbital angular momentum contribution. 
Things are subtler for spinning particles, as we now discuss. 

Considering again the case of 
a pair of spin $1/2$ emitters of momenta $p_1$ and $p_2$, the non-radiative 
amplitude reads
\begin{align}
|{\cal A}(p_1,p_2)|^2
&=\Tr\left[
(\slashed{p}_1+m)\gamma^0{\cal H}^{\dag}\gamma^0(p_1,p_2)(\slashed{p}_2-m){\cal 
H}(p_1,p_2))
\right]~.
\end{align}
When the derivatives in \cref{squared} act on ${\cal H}$ and ${\cal H}^{\dag}$ 
one recovers the 
orbital angular momentum contribution. On the other hand, the derivatives 
acting on $(\slashed{p}_1+m)$ and $(\slashed{p}_2-m)$ correspond to the spin 
contribution. This can be seen by noting that
\begin{align}
-G^{\mu\nu}_1\frac{\partial}{\partial p^{\nu}_1}(\slashed{p}_1+m)
\,\,=\,\,-\gamma^{\mu}+\frac{p_1^{\mu}}{p_1\cdot k}\slashed{k}
\,\,=\,\,
\frac{\slashed{k}\gamma^{\mu}}{2p_1\cdot k}(\slashed{p}_1+m)
+(\slashed{p}_1+m)\frac{\gamma^{\mu}\slashed{k}}{2p_1\cdot k}~,
\label{ident}
\end{align}
where in the second equality we neglected terms proportional to $k^{\mu}$ 
which 
do not contribute at the squared amplitude level. 
The r.h.s. of \cref{ident} is exactly the term that 
one considers in the interference between the first and the third  terms in 
\cref{ext+int}. In Feynman gauge, this interference term reads
\begin{align}
\Tr\left[
(\slashed{p}_2-m){\cal H}(p_1,p_2)
\left(
\frac{\slashed{k}\gamma_{\mu}}{p_1\cdot k}(\slashed{p_1}+m)
+(\slashed{p_1}+m)\frac{\gamma_{\mu}\slashed{k}}{p_1\cdot k}
\right)
{\cal H}^{\dag}(p_1,p_2))
\right]
\sum_i\eta_i\frac{p_i^{\mu}}{p_i\cdot k}
~.\label{interf}
\end{align}
The remaining interference terms resulting from \cref{ext+int} follow 
analogously. 
The spin and the orbital contributions in \cref{ext+int} combine then 
into a derivative of the non-radiative squared amplitude. Hence, 
\cref{squared} 
holds also for spinning particles. 

%
%

\section{LBK theorem with shifted kinematics}
\label{sec:shifts}

In the previous section we focused on the case with two charged particles. We 
note however that the above arguments can be straightforwardly generalized 
to an arbitrary number of external charged particles. The amplitude 
level expression in \cref{ext+int} becomes then quite cumbersome when many 
external 
charged particles (possibly of different spin) are present. However, after 
squaring the 
amplitude and summing over the polarizations one obtains an 
expression analogous to 
\cref{squared} which is relatively compact when compared to \cref{ext+int}.  
The resulting formula for $|{\cal A}(p_1,\dots,p_n,k)|^2$ 
will contain multiple derivatives acting on the non-radiative amplitude. This 
has some drawbacks, as we are going to discuss.

First of all, both the non-radiative amplitude ${\cal A}(p_1, \dots, p_n)$ 
and the radiative amplitude ${\cal A}(p_1,\dots,p_n,k)$ 
in \cref{squared} depend 
on the 
hard momenta $p_1$, \dots, $p_n$ which satisfy $\sum_i p_i=-k$. Hence, momentum 
conservation is violated in ${\cal A}(p_1,\dots,p_n)$ for finite $k$, as 
originally observed 
by Burnett and Kroll \cite{Burnett:1967km} \footnote{See also 
	\cite{Gervais:2017yxv} and the recent Refs. \cite{Lebiedowicz:2021byo, 
		Engel:2021ccn}.}. Of course, 
this 
violation is under control, since ${\cal A}(p_1,\dots,p_n)$ multiplies an NLP 
term. Thus it is an NNLP effect, 
beyond the range of validity of the LBK 
theorem. Still, this is a feature that one would like to avoid in a numerical 
implementation of \cref{squared}, since the non-radiative \emph{cross-section} 
is  typically 
generated separately from the radiative one by Monte Carlo codes.
Also, derivatives in \cref{squared} act on the squared amplitude, and not
on
the full cross-section. 
Moreover, when the number of external particles grows one is forced to take 
many derivatives of the non-radiative amplitude. This aspect might lead to 
difficulties in numerical implementations.

For these reasons, it is desirable to have another analytic formula for 
\cref{squared}. One possibility has 
been investigated in  
\cite{DelDuca:2017twk} in the context of soft gluon factorization for processes 
with at most two 
radiative particles. 
There, it is shown that for $m=0$ 
 one can recast the amplitude squared of \cref{squared} 
 in terms of 
shifted momenta as
\begin{align}
|{\cal A}(p_1,p_2,k)|^2
&=\left(\sum_{i,j=1}^2 -\eta_i\eta_j\frac{p_i\cdot p_j}{p_i\cdot k \,p_j\cdot 
	k}\right)
|{\cal 
	A}(p_1+\delta p_1,p_2+\delta p_2)|^2 
~,
\label{shift2}
\end{align}
where
\begin{align}
\delta p_1^{\mu}= \frac{1}{2} \left(-\frac{p_2\cdot k}{p_1\cdot p_2}p_1^{\mu}
+\frac{p_1\cdot k}{p_1\cdot p_2}p_2^{\mu}
-k^{\mu}\right)~,\quad 
\delta p_2^{\mu}= \frac{1}{2} \left(\frac{p_2\cdot k}{p_1\cdot p_2}p_1^{\mu}
-\frac{p_1\cdot k}{p_1\cdot p_2}p_2^{\mu}
-k^{\mu}\right)~.
\label{delta12}
\end{align}
Note that the shifts $\delta p_i^{\mu}$ satisfy $\delta p_i\cdot k=-k^2/2=0$. 
Also, 
$\delta p_i\cdot 
p_i=0$. 
Moreover, $\delta p_1^{\mu}+\delta 
p_2^{\mu}=-k^{\mu}$ and 
therefore, unlike \cref{squared}, the non-radiative amplitude in \cref{shift2} 
is expressed in terms
of conserved 
momenta. 
Finally, note that the shifts in \cref{delta12}
imply that the Mandelstam variable $s=(p_1+p_2)^2$ is shifted according to 
$s\to Q^2 $, 
where $Q$ is the invariant mass of the final-state particles in the 
non-radiative process. 

This approach can be extended to include an arbitrary number $n$ of 
(massless or massive) external charged particles.  To do so, we note that the 
	derivation in \cite{DelDuca:2017twk} computed separately the contributions 
	from the orbital and spin angular 
momentum at the amplitude level (hence specifying 
 the nature of the hard emitter). 
In this way the correspondence with 
the modern language of soft theorems of \cref{next-to-soft} and \cref{SLP} is 
made manifest.
However, as the discussion in the previous section made clear,
the form of the LBK theorem in \cref{squared} does not depend on  
the spin of the external particles. Thus, 
the shifts can be determined by looking at the orbital angular momentum only in 
the scalar case. Then, a simple calculation reveals that 
one can generalize \cref{shift2} with
\begin{align}
|{\cal A}(p_1,\dots,p_n,k)|^2
&=\left(\sum_{i,j=1}^n-\eta_i\eta_j\frac{p_i\cdot p_j}{p_i\cdot k \,p_j\cdot 
	k}\right)
|{\cal 
	A}(p_1+\delta p_1,\dots,p_n+\delta p_n)|^2 
~,
\label{shiftn}
\end{align}
where the shift on the $\ell$-th momentum $p_{\ell}$ is defined as
\begin{align}
\delta p_{\ell}^{\nu}= 
\left(\sum_{i,j=1}^n\eta_i\eta_j\frac{p_i\cdot p_j}{p_i\cdot k 
	\,p_j\cdot 
	k}\right)^{-1}
\sum_{m=1}^{n}
\left(
\eta_m\eta_{\ell}\frac{(p_{m})_\mu G_{\ell}^{\mu\nu}}{p_m\cdot k}
\right)
~.
\label{deltapi}
\end{align}
Once again, note that $\sum_{i=1}^n \delta p_i^{\mu}=-k^{\mu}$ and therefore 
momentum 
is conserved in the non-radiative amplitude. 
Consequently, the kinematical invariants $s_{ij}=(p_i+p_j)^2$ are shifted 
according to
\begin{align}
s_{ij}\to
s_{ij}\left(
1-\frac{2(p_i+ p_j)\cdot k}{s_{ij}}R_{ij}
\right)
~.
\label{deltasij}
\end{align}
Here we defined
\begin{align}
R_{ij}&=
\left(\sum_{a,b=1}^n\eta_a\eta_b\frac{p_a\cdot p_b}{p_a\cdot k 
	\,p_b\cdot 
	k}\right)^{-1} 
\left(\eta_i\frac{(p_i)_{\mu}}{p_i\cdot k}+\eta_j\frac{(p_j)_{\mu}}{p_j\cdot 
	k}\right)
\sum_{c=1}^n\eta_c \frac{p_c^{\mu}}{p_c\cdot k}~.
\end{align}
Note that \cref{deltasij} for the variable $s=(p_1+p_2)^2$ reads
\begin{align}
s\to
s\left(
1-(1-z)R_{12}
\right)
~,
\label{deltas}
\end{align}
where $z=Q^2/s$ and $Q$ is the invariant mass of all
final 
states in the non-radiative process. 
 In the 
	case of two legs, the shift in \cref{deltas} reduces to the much simpler 
	$s\to s z$, which has proven to be crucial in the Mellin space approach 
to soft 
gluon resummation at NLP for processes with no colored final states  
\cite{Bahjat-Abbas:2019fqa}. In this regard, \cref{deltas} might pave the way 
to the extension of this program to processes 
with colored final states.

The squared radiative amplitude in \cref{shiftn} can be implemented in fully 
differential 
cross-sections. To do so, it is convenient to 
express the non-radiative cross-section $d\sigma_H$ in terms of the shifted 
kinematics. However, in order to compensate for shifting momenta in the flux 
factor one has to rescale the overall cross-section by the factor 
in
\cref{deltas}.  
The generalization of \cref{crossLP} including 
NLP effects of order $(\omega_k)^0$ due to the LBK theorem then reads
\begin{align}
\frac{d\sigma_{\text{NLP-tree}}}{d^3k} & =\frac{\alpha}{(2\pi)^2} 
\frac{ 1}{\omega_k} \int d^3p_3\dots\int 
d^3p_{n}
\left(\sum_{i,j=1}^{n}-\eta_i\eta_j
\frac{p_i\cdot p_j}{(p_i\cdot k)( p_j\cdot k)}\right) \notag \\
& \qquad \left(
1-(1-z)R_{12}
\right) \,
d \sigma_H(p_1+\delta{p_1},...,p_{n}+\delta{p_{n}})
~,
\label{crossNLP}
\end{align}
where the label ``tree'' is meant to stress that there are loop corrections 
(both in QCD and QED)
to this formula, as we discuss in the next section. 
Note that the soft function in \cref{crossNLP} is the same LP soft function 
appearing in 
\cref{crossLP}. Finally, note that \cref{crossNLP} can be
extended to include also neutral particles in the final state. To this end, it 
suffices to shift only the  
momenta of the charged particles, while $Q^2=sz$ becomes the 
squared invariant mass of all (charged and neutral) final states in the 
non-radiative process. 

\section{QCD corrections with radiative jet functions}
\label{sec:loop}

It is well-known that the LP soft theorem in QED with massive particles is 
universal and does not 
receive 
loop corrections. 
At NLP, on the other hand, the LBK theorem of \cref{squared} and 
\cref{crossNLP} is modified 
at higher orders in perturbation theory, both in QED and in QCD. Nonetheless, 
virtual collinear effects 
do 
not 
contribute if the photon energy 
$\omega_k\ll\frac{m^2}{Q}$, where $m$ is 
the mass of the lightest charged particle and $Q$ is the typical energy of 
the 
process. However, this region is very narrow at high energies and does not 
include the massless region, which is needed in perturbative calculations 
involving hadronic external states. The larger region 
$\frac{m^2}{Q}\le\omega_k\sim m$ (which includes the massless limit) was 
first considered by Del Duca 
\cite{DelDuca:1990gz}. In this kinematical region, the coupling of 
the soft photon
with collinear virtual particles
generates  new small scales (i.e. $p_i\cdot k$) that prevent a Taylor 
expansion in 
the soft photon energy $\omega_k$ and generate logarithmic corrections. 
The composite operators describing these emissions at all-orders are called 
\emph{radiative 
	jets}, and have been extensively studied in the recent years 
\cite{Bonocore:2015esa, Gervais:2017yxv, Moult:2019mog, 
	Laenen:2020nrt}. 

 In contrast to the soft gluon resummation program, the 
 intricate all-order structure of scattering amplitudes is 
 not necessary for the one-loop analysis of  
the photon bremsstrahlung. In particular, in order to study QCD corrections to 
the soft photon spectrum in processes such as the $q{\bar q}\gamma$ production
at LEP  
relevant for the analysis in \cite{DELPHI:2005yew, DELPHI:2010cit} or the 
leptonic decay of the Z boson in 
hadronic collisions, it suffices to consider 
the contribution 
from the simplest of these radiative jets, denoted at one-loop as $J^{\mu(1)}$.

\begin{figure}
	\centering
	\includegraphics[width=150mm]{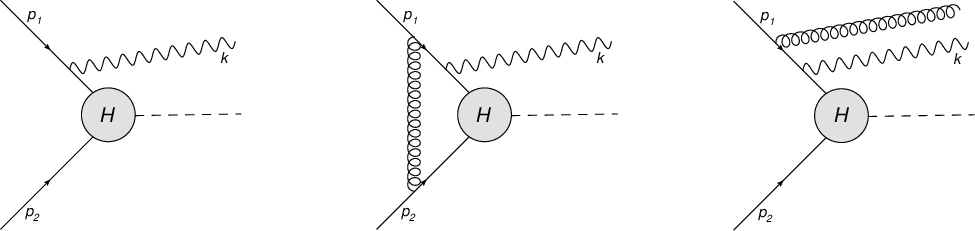}
	\caption{Sample diagrams contributing
		to the process considered in this section, i.e. the production of a 
		soft photon from the annihilation of a 
		quark-antiquark pair into a neutral state of invariant mass $Q^2$ via 
		the 
		unspecified hard subdiagram ${\cal H}(p_1,p_2)$. 
		The diagram on the left corresponds to the tree-level contribution.
		Virtual and real QCD 
		corrections are shown in the middle and the right diagrams, 
		respectively. The radiative jet function $J^{\mu(1)}$ of \cref{radjet} 
		captures the 
		contribution of the middle diagram when the momentum of the virtual 
		gluon becomes collinear to the momentum $p_1$ of the incoming quark. }
	\label{fig:loop}
\end{figure}

Specifically, let us consider a process 
with two charged massless quarks annihilating via a generic hard vertex ${\cal 
	H}(p_1,p_2)$, as shown in \cref{fig:loop}.  
In this case, the NLP soft theorems for the tree-level and the one-loop 
amplitudes read
\begin{align}
&\epsilon_{\mu}(k){\cal A}^{\mu(0)}(p_1,p_2,k)=
\left({\cal S}_{\text{LP}}^{} + {\cal S}_{\text{NLP}}^{} \right){\cal 
	A}^{(0)}(p_1,p_2)
~, \\
&\epsilon_{\mu}(k){\cal A}^{\mu(1)}(p_1,p_2,k)=
\left({\cal S}_{\text{LP}}^{} + {\cal S}_{\text{NLP}}^{} \right){\cal 
	A}^{(1)}(p_1,p_2)
+\sum_{i=1}^2 \epsilon_{\mu}(k)\,q_i\, J_i^{\mu(1)}(p_i,n,k)\,{\cal 
	A}^{(0)}(p_1,p_2)
~,
\label{NLPloop}
\end{align}
respectively. 
Here,  ${\cal S}_{\text{LP}}$ and ${\cal S}_{\text{NLP}}$ are the same as in   
\cref{SLP} and $q_i$ is the electric charge of the quark.  
Note that the upper indices $0,1$ relate to the expansion in 
$\alpha_s$ according to ${\cal 
	A}=\sum_i 
(\frac{\alpha_s}{4\pi})^i {\cal A}^{(i)}$, and correspondingly for $J^{\mu}$. 
In particular, the one-loop radiative jet function $J^{\mu(1)}$ in 
$d=4-2\epsilon$ 
dimensions reads \cite{Bonocore:2015esa} 
\begin{align}
J^{\mu(1)}(p,n,k)&=\left(\frac{\bar\mu^2}{2p\cdot k}\right)^{\e}\,
\Bigg[\left(\frac{2}{\e}+4+8\e\right)
\left(
\frac{n\cdot k}{p\cdot k}\frac{p^{\mu}}{p\cdot n}
-\frac{n^{\mu}}{p\cdot n}
\right)
-(1+2\e)\frac{ik_{\alpha}S^{\alpha\mu}}{p\cdot k}	
\notag \\	
& + \left(
\frac{1}{\e}-\frac{1}{2}
-3\e
\right)	\frac{k^{\mu}}{p\cdot k}
+(1+3\e)
\left(
\frac{\gamma^{\mu}\slashed{n}}{p\cdot n}
-\frac{p^{\mu}}{p\cdot k} \frac{\slashed{k}\slashed{n}}{p\cdot n}
\right)\Bigg]+{\cal O}(\e^2,k)~,
\label{radjet}
\end{align}
where $\bar\mu^2=4\pi\mu^2e^{-\gamma_E}$ is the $\overline{\text{MS}}$ scale  
and
$n^{\mu}$ is a light-like reference vector that in the following will be set to
$p_1$ or $p_2$, respectively.

In analogy with \cref{next-to-soft}, both ${\cal 
	S}_{\text{NLP}}$ and
$J^{\mu(1)}_i$ in \cref{NLPloop} contain 
spinor indices and therefore represent operators 
acting on the hard function as in \cref{ext+int} rather than
multiplicative factors of the non-radiative amplitude.
However, note that the dominant contribution in \cref{NLPloop} originates 
from the logarithmic terms arising from \cref{radjet}.
Thus, by isolating this logarithmic contribution, the radiative jets become a 
scalar 
factor multiplying the non-radiative amplitude.
More precisely, we obtain 
\begin{align}
\sum_{i} \epsilon_{\mu}(k)\,q_i\, J_i^{\mu(1)}
= \frac{2}{p_1\cdot p_2}
\Bigg[&
\sum_{ij}
\left(\frac{1}{\e}+
\log\left(\frac{\bar\mu^2}{2p_i\cdot k}\right)\right)
q_j\,p_i\cdot k\,
\frac{p_j\cdot\epsilon}{p_j\cdot k}
\Bigg]
+{\cal O}(\e,k^0)~.
\label{logomega}	
\end{align}

When squaring the amplitude in \cref{NLPloop} and summing over the 
 polarizations, the radiative jet 
contributes via the interference with the LP soft factor, yielding
\begin{align}
&2\Re\Bigg[
\left(\frac{p_1^{\mu} }{p_1\cdot k} - \frac{p_2 ^{\mu} }{p_2 \cdot k}  \right) 
\Bigg(
Tr[\slashed{p}_2{\cal H}^{(0)} J_{\mu}^{(1)}(p_1,p_2 ,k) \slashed{p}_1
\gamma^0{\cal 
	H}^{\dag{(0)}}\gamma^0]\notag \\
&\qquad -Tr[\slashed{p}_2J_{\mu}^{(1)}(p_2 ,p_1,k){\cal H}^{(0)} 
\slashed{p}_1
\gamma^0{\cal 
	H}^{\dag{(0)}}\gamma^0]
\Bigg)\Bigg]~.
\label{squared2}
\end{align}
Plugging \cref{radjet} into \cref{squared2} one obtains a divergent expression 
with a quite intricate structure involving various momenta and gamma matrices. 
This is the radiative jet contribution to the virtual cross-section, which  
must 
be then combined
with the real diagrams containing an unobserved real gluon (together with the 
observed real photon, as shown in \cref{fig:loop}) in order to 
cancel the soft divergences. 
After the remaining collinear singularities are absorbed in the parton 
distribution functions, one is left with a finite expression where the dominant 
term is given by ${\cal O}(\alpha_s)$  logarithmic corrections to the soft 
photon bremsstrahlung. 
The coefficient of these logarithms 
can be read directly from 
\cref{logomega}. 
In fact, the poles present 
in the other parts of the factorized cross-section cannot give any 
$\log(\omega_k)$, since this type of logarithms can only be generated by the 
collinear 
scale $p_i\cdot k$ present in the radiative jet.

 Therefore, 
	by combining \cref{NLPloop},  \cref{logomega} and \cref{squared2}
we can replace 
\cref{crossLP} with:
\begin{align}
\frac{d\sigma}{d^3k}&=
\frac{d\sigma_{\text{NLP-tree}}}{d^3k}
+\frac{\alpha_s}{4\pi}\frac{d\sigma_{\text{NLP-}J^{(1)}}}{d^3k}~,
\end{align}
where $\frac{d\sigma_{\text{NLP-tree}}}{d^3k}$ defined in \cref{crossNLP}  
contains both the standard LP contribution and the tree-level NLP correction 
and 
\begin{align}
\frac{d\sigma_{\text{NLP-}J^{(1)}}}{d^3k} &=
\frac{\alpha}{(2\pi)^2} 
\frac{ 1}{\omega_k} \int d^3p_3\dots\int 
d^3p_{n}
\left(\sum_{i=1}^{2}\eta_i
\frac{8\log\left(\frac{\bar\mu^2}{2p_i\cdot k}\right)}{p_i\cdot k}\right)
d\sigma_H(p_1,\dots,p_n)~.
\label{NLPloopfinal}
\end{align}
The result in \cref{NLPloopfinal} gives the complete logarithmic 
dependence (at order $\alpha \alpha_s$ and in the massless limit)
for processes with a single quark pair such as the leptonic decays of the $Z$ 
boson via 
quark-antiquark 
annihilation or $e^+e^-\to Z\to q 
\bar q$. The latter in particular is the process 
studied in detail by the DELPHI collaboration in \cite{DELPHI:2005yew, 
DELPHI:2010cit} where 
large deviations w.r.t. the tree-level soft spectrum have been observed.
In this regard, it is interesting to note that after setting the 
scale $\bar\mu$ to the typical high energy of the process, the QCD corrections 
in \cref{NLPloopfinal} might be large when the photon has very small energy 
and/or very low transverse momentum w.r.t. the direction of the emitting 
particle. 

It is tempting to generalize \cref{NLPloopfinal} to more generic processes, by 
simply letting the index $i$ in \cref{NLPloopfinal} run over all external 
quarks. We note however that in this case
there might be additional logarithmic corrections to 
the NLP
cross-section due to soft photon emission from 
multiple collinear 
particles connected to the hard vertex, whose computation 
is beyond the scope of this 
letter\footnote{While a field theoretical definition 
for these 
additional 
radiative jets
has been 
provided in the effective field theory language, only the relevant diagrams 
have been identified in 
the full-QCD factorization approach \cite{Larkoski:2014bxa, Gervais:2017yxv, 
	Laenen:2020nrt}. }.

Finally, it is interesting to compare 
\cref{NLPloop} with a parallel body of work that 
discussed  logarithmic corrections 
to soft theorems in QED and gravity in scattering amplitudes
\cite{Laddha:2018myi, 
	Sahoo:2018lxl}. 
Specifically, after identifying  $\log\left(\frac{\bar\mu^2}{2p_i\cdot k}\right)
\to \log\left(\omega^{-1}_k\right)$ in \cref{logomega} one recovers a similar 
structure to 
the massless limit of the results in 
\cite{Sahoo:2018lxl}\footnote{See eq.~(2.4) and eq.~(2.5) 
		there.}. However, 
note that \cite{Sahoo:2018lxl} used an unconventional prescription 
for the 
regularization of infrared divergences, while 
\cref{logomega} is expressed 
in dimensional regularization, which is more common 
in perturbative QCD calculations. 
Note also that in contrast to \cite{Sahoo:2018lxl}, where the analysis was 
completely carried in QED, here we considered a photon emission with a QCD loop 
correction.

%

\section{Discussion}
\label{sec:concl}

Measurements of soft photon spectra remain 
not understood for processes with final state  
hadrons, as the measured spectrum  
shows significant excess above the theoretical predictions based on the   
leading-power 
(LP) formula in (\cref{crossLP}).
In this letter we have considered two different corrections to 
this formula at next-to-leading power (NLP).

The first type of these corrections is of order $(\omega_k)^0\sim 1$ and it is 
due to 
the LBK theorem. We have reviewed the theorem, providing a formula 
(\cref{crossNLP}) with shifted 
kinematics which is particularly suitable for numerical implementations. 
Besides, by expressing the non-radiative amplitude in terms of conserved 
momenta, \cref{crossNLP}  underlines the 
validity of the LBK theorem at tree-level, 
which has been recently questioned in \cite{Lebiedowicz:2021byo}.

Although here we considered the photon bremsstrahlung,
	it is also worth noting that shifting the incoming momenta at NLP
	has turned out to be an efficient method for the derivation
	of threshold resummation formulae at NLP 
	in the hadroproduction of colorless final states 
	\cite{Bahjat-Abbas:2019fqa}. 
Therefore, the generalization of the shifts 
to processes with 
an arbitrary number of charged legs presented here in \cref{deltasij}  and 
\cref{crossNLP} might provide an efficient tool  
in the extension of the NLP 
 resummation to 
processes with colored final states.

The second type of corrections is of order $\log (\mu^2/p_i\cdot k)$ and it is 
due to 
QCD loop effects originating from radiative jet functions. Although radiative 
jets have been 
so far investigated mainly in the context of soft gluon resummation, they can 
be naturally applied also for the soft photon spectrum. 
In particular, in this work we have shown for the first time  
	how 
	mixed QED-QCD effects can be studied with radiative jets. 
	More specifically, we have provided a formula (\cref{NLPloopfinal})
	which gives ${\mathcal O}(\alpha_s)$ logarithmic corrections in the 
	massless limit 
	to the soft photon spectrum in processes with a single 
	quark-antiquark pair in the external states, such as the high energy limit 
	of $e^+e^-\to q \bar q \gamma$ investigated in \cite{DELPHI:2005yew, 
	DELPHI:2010cit}.
	
	The appearance of logarithmic corrections to soft 
	theorems is not surprising \cite{Bern:2014oka, 
		He:2014bga, 
		Larkoski:2014bxa, Bonocore:2014wua}. In fact, one-loop logarithmic 
		corrections 
		of the form 
	$\log(\omega_k)$ have been already 
	computed in scalar QED at the amplitude level in 
	\cite{Laddha:2018myi, Sahoo:2018lxl}, although  with a somewhat 
	unconventional 
	regularization of infrared divergences. The approach followed here with 
	a photon emission from QCD radiative jets 
	not only exploits the more common dimensional regularization to regularize 
	both collinear and soft divergences (hence making the result directly 
	implementable in partonic cross-sections such as \cref{NLPloopfinal}), but 
	emphasizes that in the 
	massless 
	limit (which is relevant at high energies) the origin of these logarithmic 
	corrections is linked to the collinear scale $p_i\cdot k$, which might be 
	small 
	not only for small energies $\omega_k$ but also for small transverse 
	momenta w.r.t. the direction of the radiating particles.

Summarizing, 
QCD corrections due to the radiative jets are a very good candidate to provide
sizable 
contributions to soft photon yields. Hence, they are important
for theoretical description of the photon radiation in processes involving 
hadrons. Specifically, they are very interesting to consider in the context 
of the observed discrepancies between theoretical predictions and experimental 
measurements for photon spectra in hadronic Z decays, as observed in 
\cite{DELPHI:2005yew, DELPHI:2010cit}. Indeed, \cref{NLPloopfinal} seems 
to go in the right direction by providing QCD corrections that (although 
suppressed in $\alpha_s$) might be particularly enhanced for very small photon 
energies and/or very low transverse momentum w.r.t. the jet axis.
However, this issue can be resolved only with a complete numerical study. Work 
in this regard is ongoing.

Finally, we note that one could reverse the problem and exploit 
QCD corrections to the LBK theorem given by radiative jet functions as a tool 
to 
obtain 
information about the 
jet structure.
A precise measurement of the soft photon spectra
probing these corrections 
would be invaluable for such studies. 
More generally, in the light of the future plans to measure ultra soft 
photons 
at LHC, we believe that there 
is much potential in the recently developed techniques for NLP corrections in 
QCD to shed new 
light on infrared photon physics.

\section*{Acknowledgments}
We thank Anton Andronic, Christian Klein-B\"osing, Peter Braun-Munzinger, 
Stefan Fl\"orchinger, Klaus Reygers and Johanna Stachel for 
stimulating discussions that led to this project. We are also grateful to 
Tim Engel, Adrian Signer, Yannick Ulrich and Leonardo 
Vernazza  for communications regarding loop 
contributions to the LBK theorem.

\bibliography{ref.bib}


\end{document}